\documentclass[useAMS,usenatbib]{mn2e}
\usepackage{times,aas_macros}
\usepackage{amssymb}
\usepackage{amsmath}
\usepackage{rotating}
\input{epsf}

\title[]
{Modelling the Extreme X-ray Spectrum of IRAS 13224-3809}

\author[C.-Y. Chiang et al.]
{Chia-Ying Chiang$^{1,2}\thanks{E-mail: cychiang@ap.smu.ca}$, D. J. Walton$^{3}$, A. C. Fabian$^{1}$, D. R. Wilkins$^{1,2}$, and L. C. Gallo$^{2}$\\
$^{1}$Institute of Astronomy, University of Cambridge, Madingley Road, Cambridge CB3 0HA\\
$^{2}$Department of Astronomy \& Physics, Saint Mary's University, 923 Robie Street, Halifax, NS B3H 3C3, Canada\\
$^{3}$Cahill Center for Astronomy \& Astrophysics, California Institute of Technology, Pasadena, CA 91125\\
}

\date{Accepted 2014 October 6.  Received 2014 October 6; in original form 2014 May 29}
\pagerange{\pageref{firstpage}--\pageref{lastpage}} \pubyear{2014}

\begin{document}

\topmargin = -0.5cm

\maketitle

\label{firstpage}

\begin{abstract}
The extreme NLS1 galaxy IRAS 13224-3809 shows significant variability,
frequency depended time lags, and strong Fe K line and Fe L features in
the long 2011 \emph{XMM-Newton} observation. In this work we study
the spectral properties of IRAS 13224-3809 in detail, and carry out a series
of analyses to probe the nature of the source, focusing in particular on the
spectral variability exhibited. The RGS spectrum shows no obvious
signatures of absorption by partially ionised material (`warm' absorbers).
We fit the 0.3-10.0 keV spectra with a model that includes relativistic
reflection from the inner accretion disc, a standard powerlaw AGN
continuum, and a low-temperature ($\sim$0.1\,keV) blackbody, which may
originate in the accretion disc, either as direct or reprocessed thermal
emission. We find that the reflection model explains the time-averaged
spectrum well, and we also undertake flux-resolved and time-resolved
spectral analyses, which provide evidence of gravitational light-bending
effects. Additionally, the temperature and flux of the blackbody component
are found to follow the $L\propto T^{4}$ relation expected for simple
thermal blackbody emission from a constant emitting area, indicating a
physical origin for this component.

%We estimated
%the black hole mass of IRAS 13224-3809 using this blackbody emission
%and obtained a mass of $\sim 10^{7} M_{\odot}$, which is consistent with
%the value derived from previous X-ray reverberation analysis.

\end{abstract}


\section{Introduction}

Some narrow-line Seyfert 1 (NLS1) galaxies display the most extreme active
galactic nucleus (AGN) properties in both spectral and timing aspects. NLS1s
are believed to harbour small black holes with high mass accretion rates
\citep{Pounds95}, and are thus frequently observed to display significant
variability. Strong soft excesses below ~2 keV \citep{Boller97} and
high-energy spectral curvature are also often seen \citep{Gallo06}. While the
soft excess can generally be modelled phenomenologically with a blackbody
component, it is likely to have an atomic origin as the required temperature
is too high for direct disc emission, and remains relatively constant over a
large range of black hole masses \citep{Gierlinski04}. Although
\citet{Gierlinski04,Gierlinski06} initially proposed that the soft excess could
potentially be produced by smeared ionised absorption, truly extreme
outflows are required to produce the smooth spectra observed, and currently
this scenario does not seem viable \citep{Schurch07}. Instead, the most likely
atomic origin is that the soft excess is produced by a series of blurred
emission lines from light elements reflected from the accretion disc (e.g.
\citealt{Crummy06, Nardini11, Walton13spin}). An alternative explanation
associates the soft excess with the Wien tail of a thermal Comptonization
component produced by the scattered disc photons \citep{Done12}. This
model can explain the spectra of a number of unobscured Type 1 AGN
\citep{Jin12}, but requires an electron population with a relatively low, fairly
constant temperature throughout the AGN population. Meanwhile, the
high-energy curvature in the spectrum can be interpreted as partial-covering
absorption or as relativistically broadened reflection from the inner disc
(e.g. \citealt{Gallo041H, Fabian04}), similar to that recently confirmed in the
Seyfert galaxy NGC\,1365 (\citealt{Risaliti13, Walton14}).

\begin{figure*}
\begin{center}
\leavevmode \epsfxsize=17.5cm \epsfbox{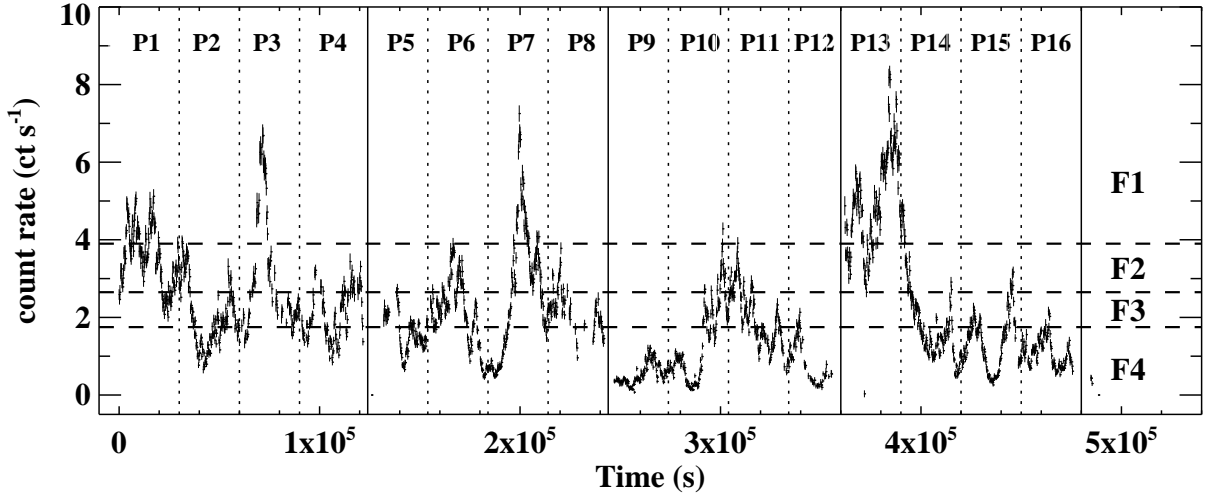}
\end{center}
\caption{The figure shows the 0.3-10 keV EPIC PN light curves with a time
bin of 200s. The dash lines in the figures show the count rate that we used
to define different flux levels (see section \ref{subsec:flux} for details). The four orbits (divided by solid vertical lines) of entire observation were separated into 16 periods, each spanning an exposure of 30 ks. Note that time intervals between each orbit were discarded.}
\label{lc_pn}
\end{figure*}

Reverberation of the soft excess was first detected in the extreme NLS1
1H\,0707-495 \citep{Fabian09, Zoghbi10}. The soft X-rays lag behind the 
hard X-rays by 30 seconds, providing strong support for a disc reflection
origin. In the standard reflection scenario, the hot corona above the
accretion disc produces the primary powerlaw continuum, which irradiates
the accretion disc and leads to the production of the reflection component.
The soft lags can be understood as the time delay between the powerlaw
(hard X-rays) and reflection (soft X-rays) components. Reverberation lags
have since been observed in a number of AGN \citep{Emmanoulopoulos11,
Zoghbi12}, and also present in stellar-mass black holes \citep{Uttley11}.
Although \citet{Legg12} claimed that the reverberation lag is a consequence
of scattering from distant (a few tens to hundreds of gravitational radii,
$R_{\rm g} = GM/c^{2}$) material, more recent work suggests this process
cannot be driving the complex lag phenomenology observed
(\citealt{Kara13feK, Walton13lag}). The work of \citet{DeMarco13} showed
that the characteristic time-scales of the reverberation lag are highly
correlated with the black hole mass, indicating that soft lags originate from
the innermost regions of the accretion flow. In addition, studies of X-ray
microlensing have shown that the X-ray emitting region of quasars are
compact \citep{Morgan12, Chartas12, Chen12}, which is consistent with the 
sizes typically inferred in the reflection scenario. Furthermore, signatures of
relativistic disc reflection have now been observed from the lensed quasar
RX\,J1131-1231 (\citealt{Reis14nat}), for which microlensing has
independently constrained the size of the X-ray emitting region to be
$R_{\rm X} \lesssim 10~R_{\rm g}$ (\citealt{Dai10}). The relativistic reflection
model offers a natural and self-consistent physical interpretation of both the
timing and spectral properties of NLS1s.  

IRAS13224-3809 (z=0.066, D $\sim2.8\times10^{8}$ pc) is a radio-quiet, extreme NLS1 which shows
many similarities with 1H0707-495, including rapid variability. The $H\beta$ line properties
(FWHM $\sim$ 650 km s$^{-1}$, flux$\sim3.5\times10^{-14}$ erg cm$^{-2}$ s$^{-1}$)
obtained from optical observations \citep{Boller93} imply a black hole mass of
$\sim 1.5\times10^{7}M_{\odot}$ \citep{Kaspi00}.
The source
was first observed with \emph{XMM-Newton} in 2002, revealing both a large
soft excess, and a sharp spectral curvature around 7-8 keV \citep{Boller03},
similar to other extreme NLS1. In particular, strong, broad iron L emission
was observed below $\sim$1\,keV \citep{Ponti10}, implying a strong
overabundance of iron, again similar to 1H0707-495 \citep{Fabian09}. In
2011, IRAS 13224-3809 was observed with \emph{XMM-Newton} for 500 ks.
The soft excess lags behind the harder continuum by $\sim$ 90\,s
\citep{Fabian13}, again providing strong support for a relativistic reflection
origin. Furthermore, a similar reverberation lag has now been detected from
the broad Fe K emission, and the lag-frequency spectrum of IRAS
13224-3809 also displays remarkable similarity to that of 1H0707-495
\citep{Kara13IRAS}. The lag analysis also implies the mass of central black hole
to be $\sim 10^{7} M_{\odot}$ (see also section \ref{subsec:mass}).
In this work, we use the latest 500 ks
\emph{XMM-Newton} observation and present detailed spectral analyses
in order to further probe the behaviour of this remarkable source.
Calculations in this paper were assumed a flat cold dark matter cosmology
with $H_{\rm 0}$ = 71 km s$^{-1}$ Mpc$^{-1}$.

\section{Data Reduction} \label{section_da}

IRAS 13224-3809 was observed with \emph{XMM-Newton} during 2011 July
19-29 (Obs. IDs 0673580101, 0673580201, 0673580301, 0673580401) for
$\sim$ 500 ks. Fig. \ref{lc_pn} shows the 0.3-10.0 keV light curves of the
European Photon Imaging Camara (EPIC) PN. The source is highly variable,
varying by a factor of $\sim$ 8 during the observation. All data have been
reduced using the Science Analysis Software ({\sevensize SAS}) version 12.0.1
with latest calibration files. The EPIC was operated in full window imaging
mode during the first observation, and in large window imaging mode in the
following three observations. We extracted source spectra of EPIC MOS and 
PN using  a circular region with a radius of 35 arcsec. Background spectra
were extracted using a region with the same size from a source-free region.
Some regions of the EPIC PN CCD are affected by the Cu-K lines from the
electric circuits behind that cause contamination on the background at 8.0
and 8.9 keV, and these regions were avoided during background selection.

\begin{figure}
\begin{center}
\leavevmode \epsfxsize=8.5cm \epsfbox{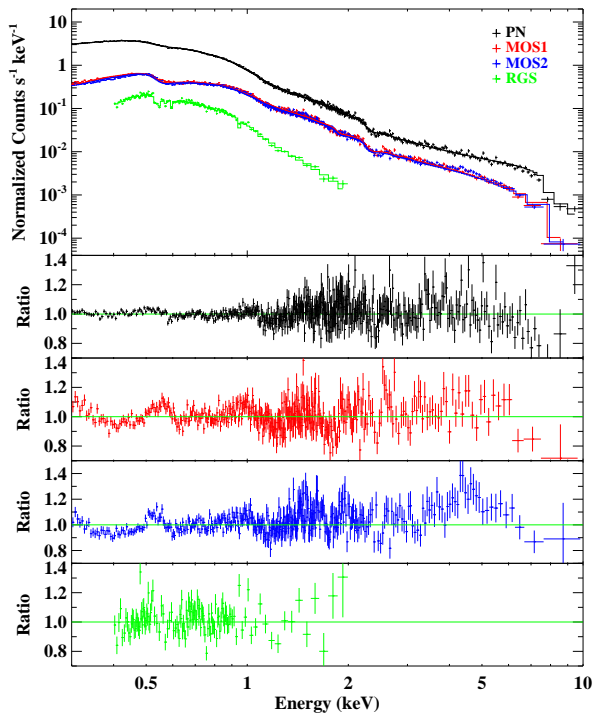}
\end{center}
\caption{The figure shows the time-averaged EPIC and RGS spectra fitted with
the best-fitting relativistic reflection model. These spectra have been re-binned 
for clarity. The PN, MOS1, MOS2, RGS data are presented using black, red, blue, 
and green points, respectively.}
\label{ta}
\end{figure}

\begin{figure}
\begin{center}
\leavevmode \epsfxsize=8.5cm \epsfbox{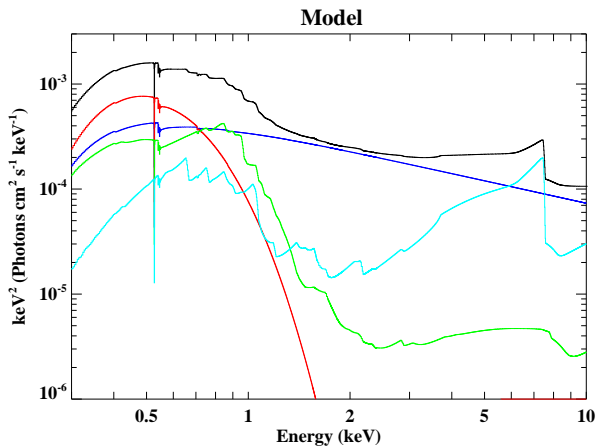}
\end{center}
\caption{The figure shows the decomposed model used for fitting. The top black line stands for the combined model.}
\label{ta}
\end{figure}

The spectrum of IRAS 13224-3809 is steep and background is important at high
energies.
Periods of high background were removed following the method outlined by
\cite{Piconcelli04}, which determines the background threshold that maximises
the signal-to-noise (S/N) in a chosen band; here we maximise the S/N for the
full 0.3--10.0\,keV EPIC bandpass. We obtained total of $\sim$ 350 ks,
$\sim$ 401 ks, $\sim$ 407 ks of good exposure time for PN, MOS1, and MOS2,
respectively. The redistribution matrix file (RMF) and the ancillary response file
(ARF) were created using the {\sevensize RMFGEN} and {\sevensize ARFGEN}
tools. We examined all the observations using the {\sevensize EPATPLOT} task
in {\sevensize SAS} and found that they are not piled up. When necessary,
spectra extracted from different orbits have been reduced separately, and then
combined using {\sevensize ADDASCASPEC}, part of the {\sevensize HEASOFT}
distribution\footnote{http://heasarc.gsfc.nasa.gov/lheasoft/}. Only spectra of
from same detector were combined. We rebinned the spectra to have a
minimum of 30 counts in each energy bin using the {\sevensize GRPPHA} tool.
In this work, we use EPIC data over the full 0.3-10.0 keV bandpass, but note
that there are few data points above $\sim$ 8 keV due to the high relative
contribution of the background in this band, a consequence of the steep source
spectrum (see \citealt{Fabian13}).

We also reduced the Reflection Grating Spectrometer (RGS) data following the
standard procedures. Both the first order spectra of RGS1 and RGS2 data were
extracted. In order to analyse the data more efficiently, we used the {\sevensize
RGSCOMBINE} tool to combine RGS1 and RGS2 spectra into a single spectrum.
We analyse the RGS spectrum over the 0.4-2.0 keV bandpass, which is broadly
well-calibrated with the EPIC spectra.

\section{Data Analysis and Comparison}
\subsection{Time-averaged Spectra} \label{sec_ta}

The initial analysis of time-averaged EPIC PN spectrum can be found in
\citet{Fabian13}, and we begin by extending this time-averaged analysis to
include the MOS and RGS data. Based on that used in \citet{Fabian13}, we
construct a model to fit the EPIC and RGS spectra composed of a powerlaw
continuum, two blurred reflection components, and a very soft blackbody 
component ({\sevensize BBODY}), which contributes at the lowest energies 
probed. The double reflection components primarily account for the 
soft excess, and explain the $\sim$ 0.8 keV Fe L feature significantly better
than the smooth curvature produced by Comptonization. Galactic absorption 
was modeled using the {\sevensize TBNEW} neutral absorption 
code\footnote{http://pulsar.sternwarte.uni-erlangen.de/wilms/research/tbabs} 
(Wilms et al., in preparation), with the solar abundances presented in
\cite{Wilms00}. We used the {\sevensize EXTENDX} grid, which is an extended
version of the {\sevensize REFLIONX} grid \citep{Ross05} that allows for a wider
range of iron abundance (0.1-20 times solar abundance), to model reflected
emission. The {\sevensize RELCONV} \citep{relconv} kernel acts on the reflected
emission to account for relativistic effects. We assumed the inner edge of the
accretion disc to extend down to the innermost stable circular orbit (ISCO), and
the outer radius to be 400 $R_{g}$. Limb-darkening effects have been
considered. The model is expressed as {\sevensize TBNEW*(POWERLAW +
BBODY + RELCONV*(EXTENDX$_{1}$ + EXTENDX$_{2}$))}, and fits the data well
(see Fig. \ref{ta} and Table \ref{ta_ta}). The parameters obtained are generally
consistent with those presented in \citet{Fabian13}, though the Galactic
absorption column and iron abundance are slightly higher. These parameters
are sensitive in the soft energy bands and cross-calibration differences
between PN, MOS and RGS, the latter two now being included, may lead to
mildly different best-fitting values.

\begin{table*}
 \caption{The table lists parameters of the best-fitting reflection model of EPIC 
and RGS spectra, in which $N_{\mbox{\scriptsize H}}$ is given in $10^{20}$
cm$^{-2}$, and $\xi$ in erg cm s$^{-1}$. The normalisation of the powerlaw
component is expressed in photons keV$^{-1}$ cm$^{-2}$ s$^{-1}$, and that 
of the blackbody component in $10^{37}$ erg s$^{-1}$ kpc$^{-2}$.  $F_{\rm BB}$ were calculated over the 0.01-10.0
keV band, while flux of other components were measured between 0.1
and 100 keV. The hard upper limit of $q_{1}$
is 10, and that of $A_{\rm Fe}$ is 20.  The reflection fraction $\mathcal{R}$ is defined as (flux of reflector)/($F_{\rm RDC}$+$F_{\rm PLC}$).}
\label{ta_ta}
\centering
\begin{tabular}{@{}llcc}
\hline\hline
Component & Parameter & powerlaw emissivity & broken powerlaw emissivity\\
\hline
TBNEW & Absorption column, $N_{\rm H}$  & $6.3\pm0.1$ & $6.1^{+0.3}_{-0.1}$\\
BBODY & Temperature, $kT$ ($10^{-2}$ keV) & $9.4\pm0.1$ & $9.3^{+0.1}_{-0.2}$\\
 & Norm & $(4.0\pm0.1)\times10^{-5}$ & $3.6^{+0.3}_{-0.2}\times10^{-5}$\\
 & $F_{\rm BB}$ ($10^{-13}$ erg cm$^{-2}$ s$^{-1}$) & $33.4\pm0.4$ & $29.9^{+2.4}_{-1.2}$\\
POWERLAW & Photon index, $\Gamma$ & $2.67^{+0.03}_{-0.01}$ & $2.71\pm0.02$\\
 & Norm & $(3.7\pm0.1)\times10^{-4}$  & $(3.8\pm0.1)\times10^{-4}$\\
 & $F_{\rm PLC}$ ($10^{-13}$ erg cm$^{-2}$ s$^{-1}$) & $41.7\pm0.3$ & $43.3^{+1.2}_{-1.1}$\\
RELCONV & Inner emissivity index, $q_{1}$ & $4.3^{+0.3}_{-0.1}$ & $>9$\\
 & Outer emissivity index, $q_{2}$ & - & $3.4^{+0.3}_{-0.2}$\\
 & Spin parameter, $a^{*}$ & $>0.989$ & $0.990^{+0.001}_{-0.003}$\\
 & $R_{\rm break}$ ($R_{\rm g}$) & - & $2.1\pm0.1$\\
 & Inclination, $i$ (deg) & $64.9\pm0.3$ & $64.6^{+0.6}_{-0.7}$\\
EXTENDX & Iron abundance /solar, $A_{\rm Fe}$ & $18.1\pm0.7$ & $> 18.1$\\
 & Ionization parameter, $\xi_{1}$ & $499^{+3}_{-12}$  & $498^{+4}_{-38}$\\
 & ${\rm Norm}_{1}$ & $(1.7\pm0.1)\times10^{-8}$ & $2.2^{+0.1}_{-0.3}\times10^{-8}$\\
 & $\mathcal{R}_{1}$ & $\sim0.32$ & $\sim0.33$\\
 & Ionization parameter, $\xi_{2}$ & $20.0^{+0.2}_{-0.6}$ & $20.5^{+0.3}_{-1.1}$\\
 & ${\rm Norm}_{2}$ & $4.6^{+0.8}_{-0.2}\times10^{-6}$ & $5.9^{+0.3}_{-0.9}\times10^{-6}$\\
 & $\mathcal{R}_{2}$ & $\sim0.14$ & $\sim0.15$ \\
 & $F_{\rm RDC}$ ($10^{-13}$ erg cm$^{-2}$ s$^{-1}$) & $36.5^{+0.5}_{-0.8}$ & $41.0^{+1.3}_{-3.7}$\\
$\chi^{2}/d.o.f.$ & & 3790/2890 &  3633/2892\\
\hline\hline\\
\end{tabular}
\end{table*}

The best-fitting parameters again show that IRAS 13224-3809 harbours a
rapidly-spinning central black hole. The emissivity profile, i.e. the radial
illumination of the accretion disc, is first assumed to have a powerlaw form, following
$\epsilon(r) \propto r^{-q}$. We then tested a more sophisticated model with
a broken powerlaw emissivity profile, which follows $\epsilon(r) \propto r^{-q_{1}}$ 
within some break radius $R_{\rm break}$, and $\epsilon(r) \propto r^{-q_{2}}$ 
outside that radius. We list results obtained from both models
in Table \ref{ta_ta} for comparison. 
The
emissivity indices are similar to those presented in \citet{Fabian13} as well.
The steep inner emissivity index $q_{1}$ implies strong gravitational
light-bending effects near the central black hole. There are some residuals
around $\sim$ 0.5 keV and $\sim$ 1.4 keV in the EPIC spectra, which might
be hints of absorption. However, these features are not present in the RGS
spectrum, and no additional absorption components are required to model the
data. The RGS spectrum does not show any distinct feature, and we conclude
that the spectra of IRAS 13224-3809 are not modified by warm absorbers.

Analysis of the time-averaged spectra reveals the general spectral properties of
IRAS 13224-3809. However, during the $\sim$ 500 ks observation, the source
flux varies substantially. In order to investigate any spectral variability
associated with this flux variability, we proceed to carry out a flux-resolved
spectral analysis.

\subsection{Flux-resolved Spectra} \label{subsec:flux}

\begin{figure}
\begin{center}
\leavevmode \epsfxsize=8.5cm \epsfbox{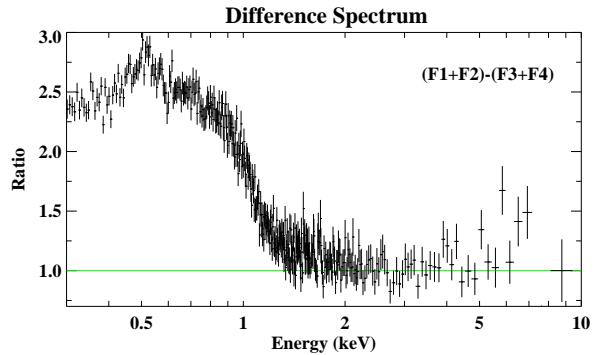}
\end{center}
\caption{The figure shows the EPIC PN difference spectrum fitted with an
absorbed powerlaw. A soft excess can be seen below $\sim$ 1 keV}
\label{diff_pow}
\end{figure}

We divided the EPIC PN 0.3-10 keV light curves into four flux intervals, and
extracted a spectrum from each. The dashed lines in Fig. \ref{lc_pn} show the
counts that have been chosen to develop the flux intervals. Each flux bin has a
similar number of total counts ($\sim1.9\times10^{5}$). The same technique
has been also used to investigate the spectral variability in MCG--6-30-15 
\citep{Vaughan04,Chiang11}. We first examined the difference spectrum
between the two highest flux states and the two lowest, i.e. the result of
subtracting the low-flux (F3+F4) spectrum from the high-flux (F1+F2)
spectrum. We modeled the difference spectrum with an absorbed powerlaw
and found a clear soft excess below $\sim$ 1.5 keV (Fig. \ref{diff_pow}). The
soft blackbody component of IRAS 13224-3809 may vary and cause the 
difference in the soft band between high and low-flux states. Alternatively, 
changes in the ionisation state of the blurred reflection component may also 
result in difference in the low-energy band. We found that the soft excess in
the difference spectrum can be fitted either by including a blackbody
component of $kT\sim0.12$ keV or a reflection component, and further
analysis is required to probe the real origin of this feature.

\begin{figure}
\begin{center}
\leavevmode \epsfxsize=8.5cm \epsfbox{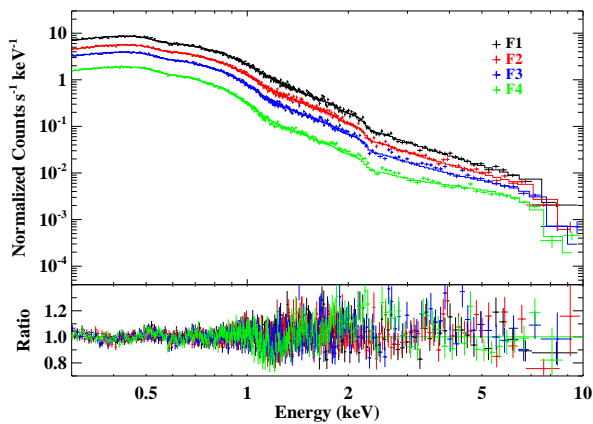}
\end{center}
\caption{The figure shows the EPIC PN flux-resolved spectra fitted with the
relativistic reflection model. Spectra of different flux states are shown using data
points of different colours. These spectra have been re-binned for clarity.}
\label{flux}
\end{figure}

\begin{table*}
 \caption{The table lists parameters of the best-fitting model of the flux-resolved
spectra, where $N_{\mbox{\scriptsize H}}$ is also given in $10^{20}$ cm$^{-2}$,
and $\xi$ in erg cm s$^{-1}$. Fluxes are again given in $10^{-13}$ erg
cm$^{-2}$ s$^{-1}$, and $\xi$ in erg cm s$^{-1}$. If parameters are bound over
the spectra, only values of F1 are listed. Since $\Gamma$ was restricted in a
narrow range and easily to hit the hard limit when calculating uncertainties, we
only list the best-fitting values.}
\label{ta_flux}
\centering
\begin{tabular}{@{}llcccc}
\hline\hline
Component & Parameter & \multicolumn{4}{c}{Value}\\
 & & F1 & F2 & F3 & F4\\
\hline
TBNEW & $N_{\rm H}$  & $6.9^{+0.4}_{-0.2}$ & - & - & -\\
BBODY & $kT$ ($10^{-2}$ keV) & $9.6^{+0.2}_{-0.3}$ & $9.3^{+0.3}_{-0.5}$ & $8.8^{+0.2}_{-0.3}$ & $8.3\pm0.2$\\
 & Norm & $7.4^{+0.6}_{-0.4}\times10^{-5}$ & $5.8^{+0.7}_{-0.5}\times10^{-5}$ & $(5.6\pm0.3)\times10^{-5}$ & $(3.1\pm0.2)\times10^{-5}$ \\
 & $F_{\rm BB}$ & $62.0$ & $48.9$ & $46.6$ & $25.8$ \\
POWERLAW & $\Gamma$ & 2.80 & 2.76 & 2.63 & 2.63 \\
 & Norm & $(1.2\pm0.1)\times10^{-3}$ & $7.0^{+0.2}_{-0.3}\times10^{-4}$ & $(3.6\pm0.1)\times10^{-4}$ & $(1.2\pm0.1)\times10^{-4}$ \\
 & $F_{\rm PLC}$ & $151.5$ & $84.3$ & $38.7$ & $13.1$ \\
RELCONV & $q$ & $3.2^{+0.6}_{-0.5}$ & $3.8^{+0.4}_{-0.6}$ & $4.1\pm0.4$ & $5.3^{+0.3}_{-0.4}$ \\
% & $a^{*}$ & ($0.990$) & - & - & - \\
% & $i$ (deg) & ($64.6$) & - & - & - \\
EXTENDX & $A_{\rm Fe}$ & $16.9^{+2.0}_{-1.1}$ & - & - & - \\
 & $\xi_{1}$ & $990^{+70}_{-210}$ & $500^{+90}_{-160}$ & $260^{+300}_{-40}$ & $220^{+80}_{-40}$ \\
 & ${\rm Norm}_{1}$ & $2.0^{+1.1}_{-0.3}\times10^{-8}$ & $3.2^{+2.9}_{-0.6}\times10^{-8}$ & $4.3^{+1.2}_{-2.3}\times10^{-8}$ & $3.5^{+2.2}_{-1.9}\times10^{-8}$ \\
 & $\mathcal{R}_{1}$ & $\sim0.32$ & $\sim0.33$ & $\sim0.36$ & $\sim0.40$\\
 & $\xi_{2}$ & $20^{+1}_{-2}$ & - & - & - \\
 & ${\rm Norm}_{2}$ & $6.9^{+1.4}_{-1.7}\times10^{-6}$ & $6.4^{+2.7}_{-2.1}\times10^{-6}$ & $3.8^{+1.1}_{-0.4}\times10^{-6}$ & $3.6^{+0.4}_{-0.3}\times10^{-6}$ \\
 & $\mathcal{R}_{2}$ & $\sim0.07$ & $\sim0.10$ & $\sim0.14$ & $\sim0.27$\\
 & $F_{\rm RDC}$ & $97.8$ & $63.0$ & $39.3$ & $22.6$ \\
$\chi^{2}/d.o.f.$ &  & \multicolumn{4}{c}{2214/2085}\\
\hline\hline\\
\end{tabular}
\end{table*}

We simultaneously modeled the four flux-resolved spectra with the same model 
used for the time-averaged spectra in section \ref{sec_ta}. Since the exposure
of  each flux-resolved spectra is obviously substantially shorter than the
time-averaged spectra, we use a simpler emissivity profile consisting of a single 
powerlaw to improve the efficiency of fitting and better constrain the parameters. 
Parameters which should remain constant on observational timescales, i.e. the 
black hole spin $a^{*}$ and the disc inclination angle $i$, have been fixed to the 
same values obtained in time-averaged analysis (Table \ref{ta_ta}). Since for this 
analysis we focus on the PN data only, the Galactic absorption column $N_{\rm 
H}$ and iron abundance $A_{\rm Fe}$ might be mildly different from those 
obtained from the combined fit to the PN, MOS and RGS time-averaged spectra. 
These parameters were allowed to vary overall, but were linked across each of
the flux bins. We also linked the the ionisation parameter of the second reflection component 
($\xi_{2}$, the lower one) as it showed no distinct evolution between different flux states.

The photon index of the powerlaw component $\Gamma$ has been found to
vary by a few per cent with flux states in some NLS1s. In Fig. \ref{flux} it can be 
seen that the shape of each spectrum is fairly similar, which implies that
photon index does not vary substantially between the different flux states. To
confirm this, we fit the flux-resolved spectra between 1-2 keV, which is an 
energy band free of absorption and dominated by the powerlaw continuum, with 
a single powerlaw and find that $\Gamma$ varies in a small range ($\sim$ 3 per 
cent). We therefore restrict the photon index to be in the range of 2.63-2.8 (the 
best-fitting value $\Gamma=2.71$ from the time-averaged analysis $\pm$3\%). 
Previous analyses of 1H0707-495 also found that the photon index varies within a 
small range (2.95-3.05) across different states.

The rest of the parameters in the model are allowed to vary between the
different flux states, and the results are presented in Table \ref{ta_flux}. As 
shown in Fig. \ref{flux}, the model works well on the flux-resolved spectra. It
can clearly be seen that the photon index tends to be softer during high-flux 
states, and the normalisation of the powerlaw component decreases from
high-flux states to low-flux states. The flux of reflection components seems to 
follow the same trend of the powerlaw component. The ionisation parameter of 
the first reflection component appears to be proportional to the flux of the 
powerlaw component. The changing ionisation parameter might be the cause
of the soft excess present in the difference spectrum (Fig. \ref{diff_pow}). 
However, the blackbody component also appears to vary between different flux 
states and also contributes to the difference spectrum. The emissivity index 
appears to be steeper when fluxes are lower, implying more emission comes
from the inner disc when the source is fainter.

%\begin{figure*}
%\begin{center}
%\includegraphics[scale=0.5]{lc_16periods.eps}
%\end{center}
%\caption{The figure shows how the entire observation was separated into 16 periods, each spanning an exposure of 30 ks. The time bin is 200 s.} \label{lc_16}
%\end{figure*}

\subsection{Time-resolved Spectra} 
\subsubsection{30 ks slices} \label{subsubsec:30ks}

In order to further probe the general trends of the source, we extracted a
series of time-resolved spectra from each of the EPIC detectors, similar to
recent analyses by \cite{Reis12, Marinucci14, Walton14}. Each interval spans
a duration of 30 ks (see Fig. \ref{lc_pn}), giving 16 intervals in total. Note
that the actual good time varies depending on the behaviour of the
background. For each interval, we independently applied the model used 
previously in sections \ref{sec_ta} and \ref{subsec:flux}. Similar to our
flux-resolved spectal analysis, parameters that should not vary on
observational timescales were fixed to the best-fitting values from our
time-averaged analysis (including now the iron abundance and the neutral
absorption column, given the consistent set of detectors used for these
analyses; see Table \ref{ta_ta}), and the photon index was again restricted to
lie between $\Gamma=2.63-2.8$. These 16 periods cover a wide variety of
flux states, providing additional data over a broad dynamic range with which
to further examine the parameter trends obtained with our flux-resolved
spectra. In Table \ref{ta_tr} we list the best-fitting parameters obtained, along
with the absorption-corrected 0.1-100.0 keV fluxes of the powerlaw
continuum and the total reflected emission ($F_{\rm PLC}$ and $F_{\rm RDC}$ 
respectively), and the 0.01-10.0 keV absorption-corrected fluxes of the
blackbody component ($F_{\rm BB}$). All fluxes were calculated using the
{\sevensize CFLUX} model in {\sevensize XSPEC}. 

\begin{figure*}
    \begin{center}
        \begin{minipage}{1\textwidth}
            \begin{minipage}{0.5\textwidth}
                \begin{center}
                    \leavevmode \epsfxsize=8.5cm {\epsfbox{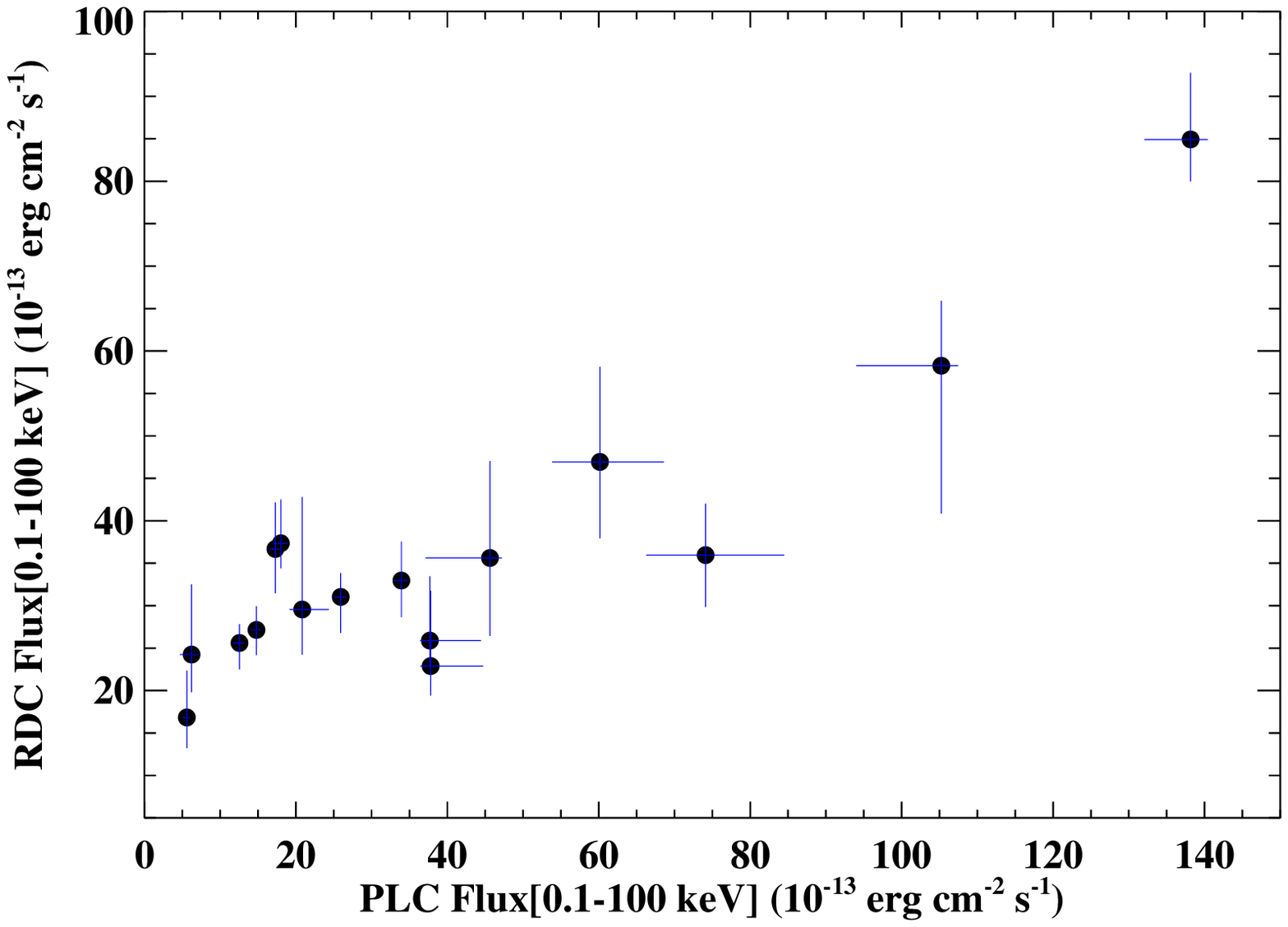}}
                    (a)
                \end{center}
            \end{minipage}
            \begin{minipage}{0.5\textwidth}
                \begin{center}
                     \leavevmode \epsfxsize=8.5cm {\epsfbox{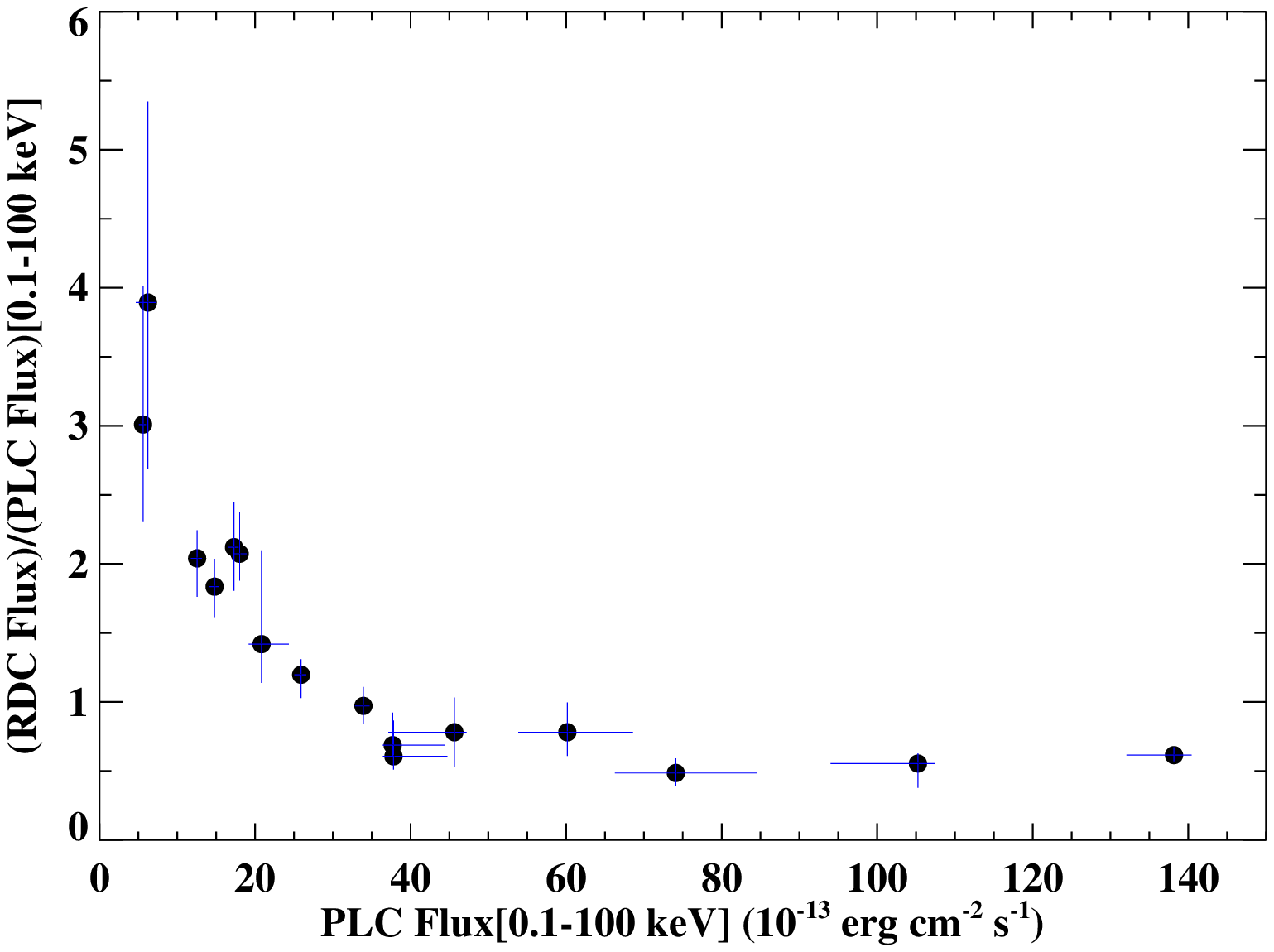}}
                     (b)
                \end{center}
            \end{minipage}
        \end{minipage}
        \begin{minipage}{1\textwidth}
            \begin{minipage}{0.5\textwidth}
                \begin{center}
                    \leavevmode \epsfxsize=8.5cm {\epsfbox{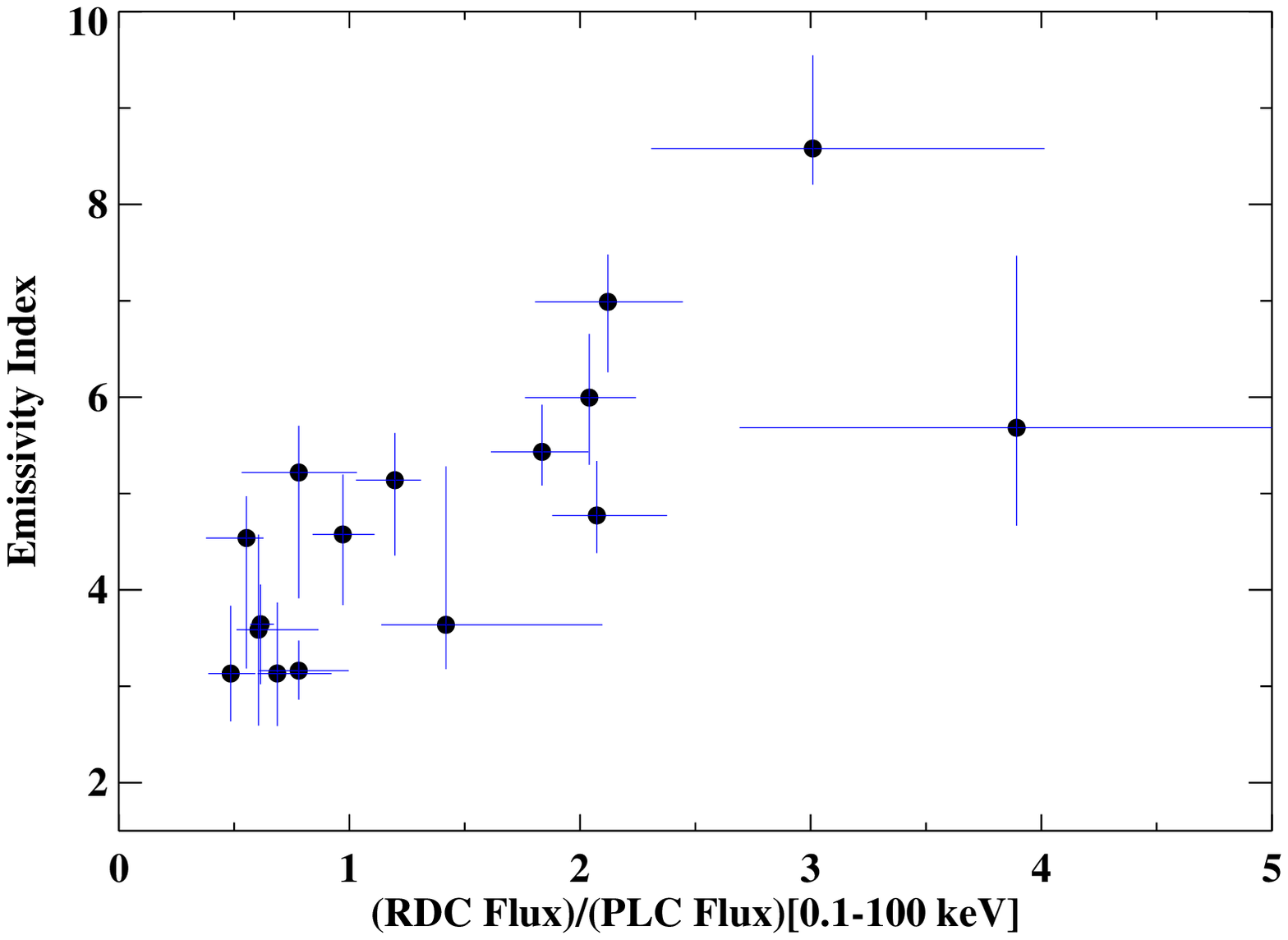}}
                    (c)
                \end{center}
            \end{minipage}
            \begin{minipage}{0.5\textwidth}
                \begin{center}
                     \leavevmode \epsfxsize=8.5cm {\epsfbox{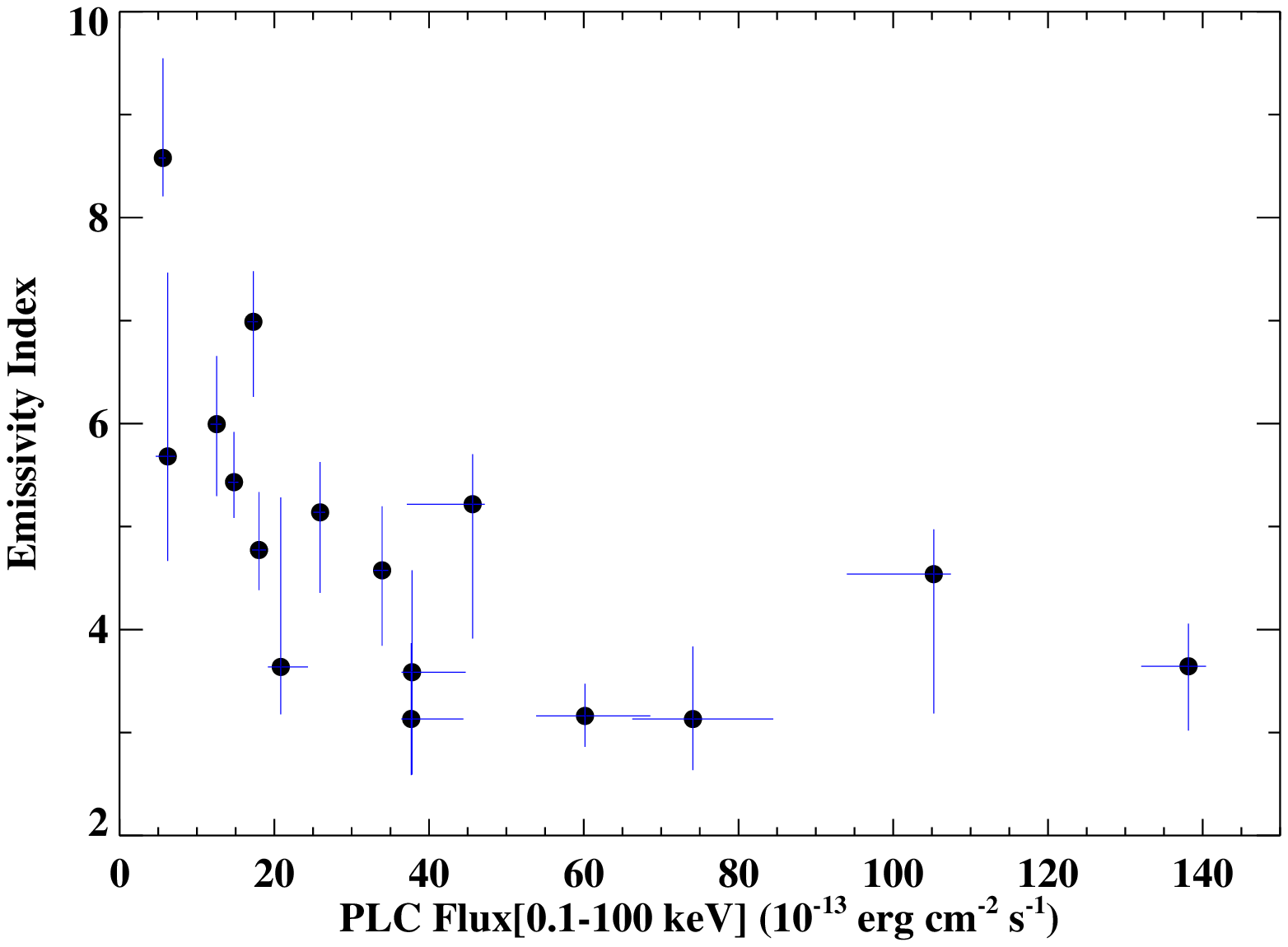}}
                     (d)
                \end{center}
            \end{minipage}
        \end{minipage}
        \begin{minipage}{1\textwidth}
            \begin{minipage}{0.5\textwidth}
                \begin{center}
                    \leavevmode \epsfxsize=8.5cm {\epsfbox{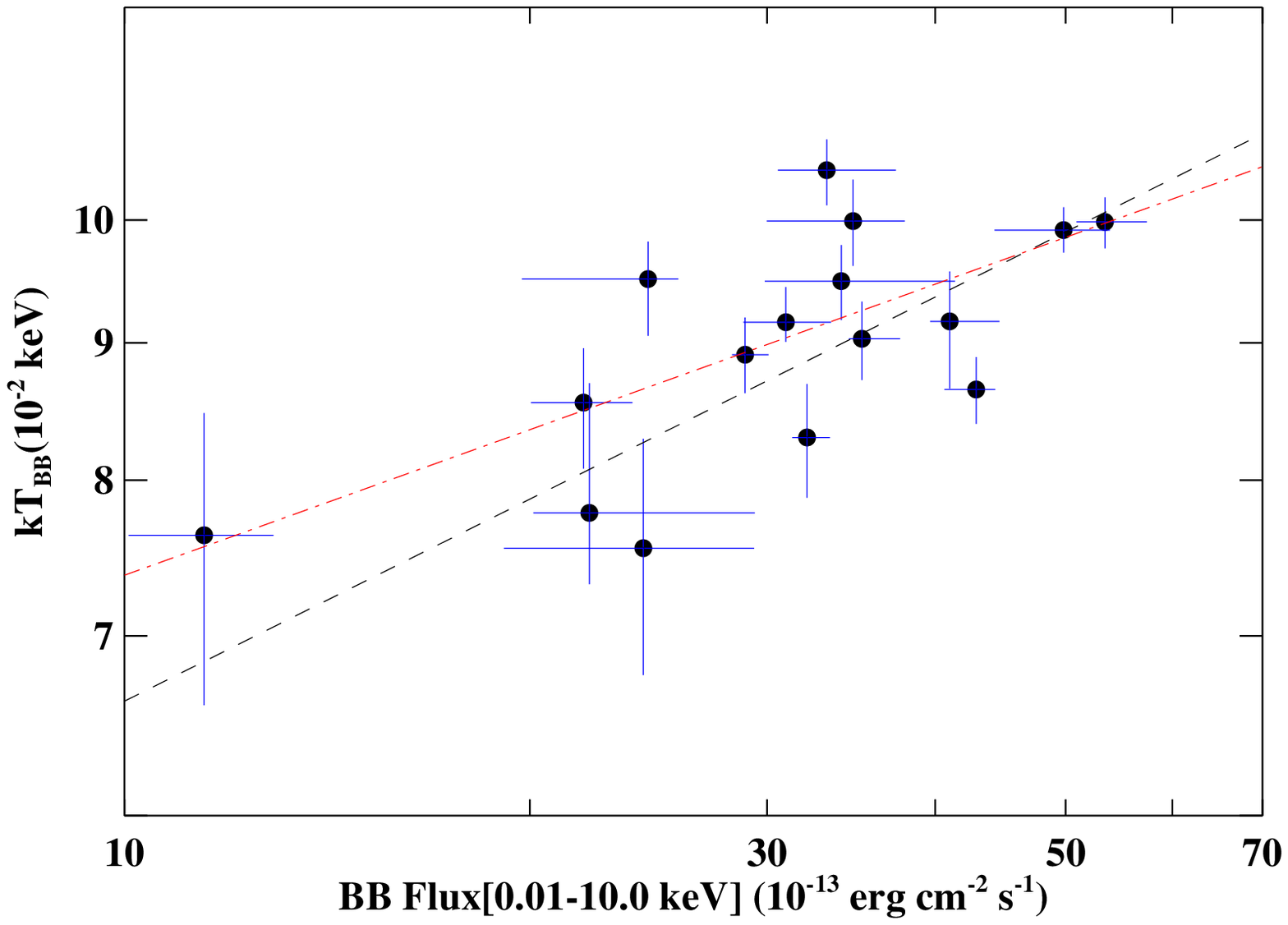}}
                    (e)
                \end{center}
            \end{minipage}
            \begin{minipage}{0.5\textwidth}
                \begin{center}
                     \leavevmode \epsfxsize=8.5cm {\epsfbox{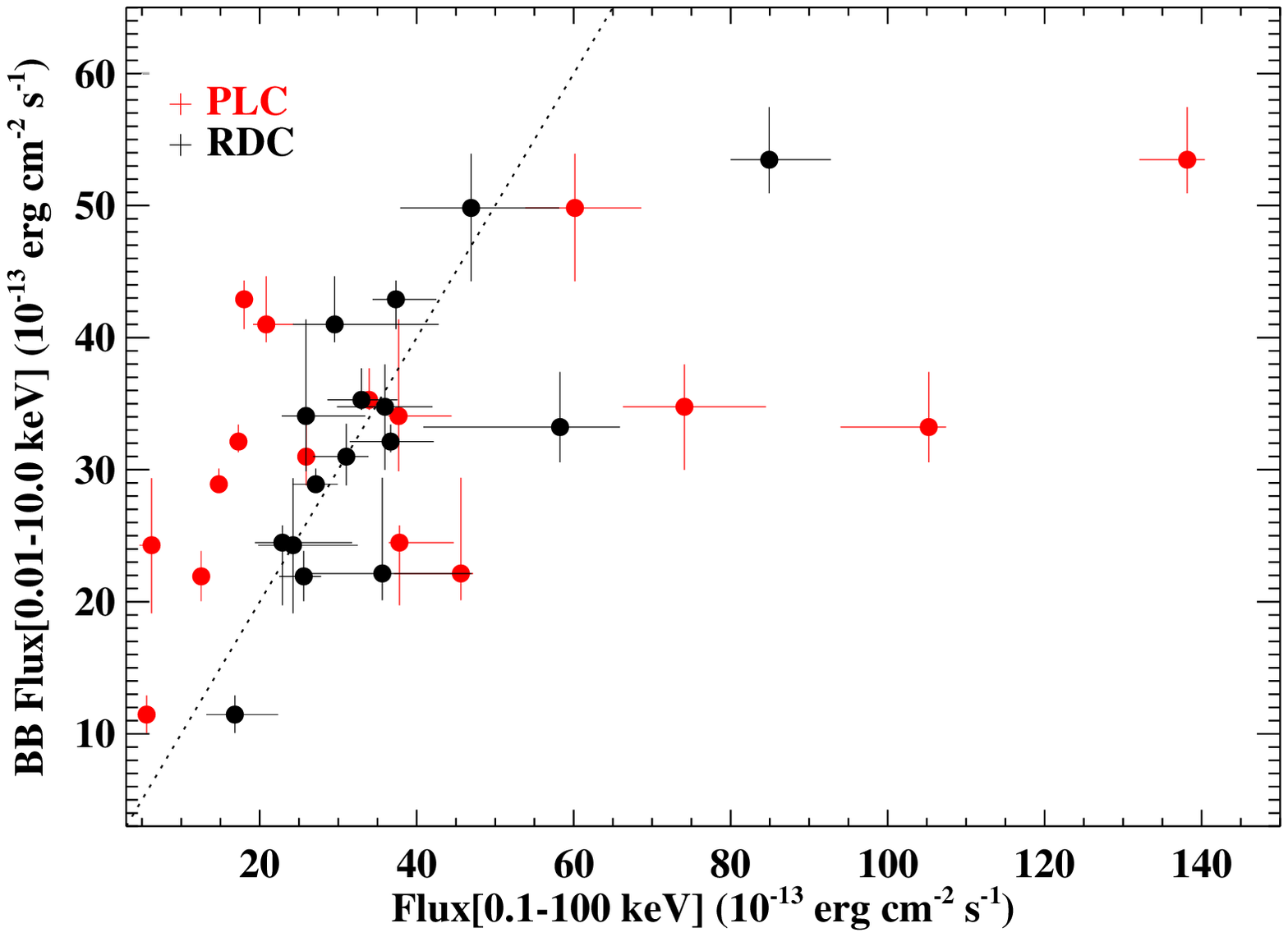}}\\
                     (f)  
                \end{center}
            \end{minipage}
            \caption{These figures show general trends of IRAS 13224-3809. PLC and RDC state for powerlaw component and reflection-dominated component, respectively. Fluxes were calculated over the 0.01-10.0 keV energy band for $F_{\rm BB}$, and over the 0.1-100 keV energy band for $F_{\rm PLC}$ and $F_{\rm RDC}$. The black dash line in figure (e) stands the $L\propto T^4$ relation and red dot dash line is drawn using the best-fitting slope ($0.18\pm0.04$) of the data. The black dot line in figure (f) shows the expected relation when either $F_{\rm PLC}\propto F_{\rm BB}$ or $F_{\rm RDC}\propto F_{\rm BB}$ follows.}
            \label{parameter}
        \end{minipage}
    \end{center}
\end{figure*}

Fig. \ref{parameter} shows a series of plots examining the observed evolution
for several of the key model parameters. It can been seen in Fig.
\ref{parameter}a that the reflected flux correlates strongly with the powerlaw
flux, although we note that the powerlaw flux varies over a larger range than
the reflected flux does. Panels b-d show that the source becomes progressively
reflection-dominated when it is fainter, and that the emissivity index also rises
as the observed flux drops and the source becomes more reflection-dominated, 
consistent with the conclusions from the flux-resolved analysis.

In panels e and f we examine the behaviour of the blackbody component. First,
we investigate the evolution of the temperature and luminosity of this
component, and find that these parameters seem to broadly follow the $L
\propto T^{4}$ relation expected for blackbody emission from a constant 
emitting area (black dash line in Fig. \ref{parameter}e). We first statistically 
confirm the correlation between these quantities by conducting a simple 
Spearman's rank correlation test and found a Pearson correlation coefficient of 
$\sim$ 0.58, corresponding to a false-alarm probability of $\sim$ 0.0093. We 
then used the IDL routine {\sevensize 
MPFITEXY}\footnote{http://purl.org/mike/mpfitexy}, an which relies
in turn on the {\sevensize MPFIT} package \citep{mpfit}, to fit the data
accounting for the uncertainties in both the luminosity and the temperature,
and found a slope of $0.18\pm0.04$ (red dot dash line in Fig. \ref{parameter}e),
which is close to the expected value 0.25. The good agreement with the basic
expectation for thermal emission supports a physical origin for the blackbody 
component. This may originate through heating of the inner accretion disc
by the powerlaw continuum and/or the reflected emission; a more detailed
discussion is given in section \ref{sec_bb}. We stress that although the blackbody 
component dominates in the observed bandpass of the soft excess, the broad
iron L feature (see \citealt{Fabian13}), is accounted for by the reflected 
emission.
%We stress again though that the 
%majority of the soft excess in the observed bandpass, and in particular the
%broad iron L feature (see \citealt{Fabian13}), is accounted for by the reflected 
%emission. 
In Fig. \ref{parameter}f we plot fluxes of the powerlaw continuum and
the reflected emission against the blackbody flux. It seems that $F_{\rm BB}$ is
correlated with both $F_{\rm PLC}$ and $F_{\rm RDC}$. The Pearson correlation
coefficients  are $\sim$ 0.51 for $F_{\rm PLC}$ and $F_{\rm BB}$, and $\sim$
0.69 for $F_{\rm RDC}$ and $F_{\rm BB}$. 

Finally, while the results presented in Table \ref{ta_tr} suggest the ionisation
parameter of the first (higher  ionisation) reflection component may correlate
with the powerlaw flux, ultimately the ionisation of this component is not well
constrained in our time-resolved analysis. Otherwise, these results help to 
confirm the trends found in the flux-resolved analysis, shedding further light
on the extreme nature of this source.

\begin{table*}
 \caption{The table lists parameters of different periods (see Fig. \ref{lc_pn}) of
data. The absorbed (absorption-corrected) fluxes of the blackbody $F_{\rm BB}$,
powerlaw $F_{\rm PLC}$ and reflection components $F_{\rm RDC}$ are given in
$10^{-13}$ erg cm$^{-2}$. $F_{\rm BB}$ were calculated over the 0.01-10.0
keV band, while $F_{\rm PLC}$ and $F_{\rm RDC}$ were measured between 0.1
and 100 keV. Temperatures of the blackbody component are given in 10$^{-2}$
keV, and $q$ states the emissivity index. Again we only list the best-fitting
values of $\Gamma$ here.}
\label{ta_tr}
\centering
\begin{tabular}{@{}cccccccccccc}
\hline\hline
Period & $\Gamma$ & $kT_{\rm BB}$ & $q$ & $\xi_{\rm 1}$ & $F_{\rm BB}$ & $F_{\rm PLC}$ & $F_{\rm RDC}$  & $\mathcal{R}_{1}$ & $\mathcal{R}_{2}$ & $\chi^{2}_{\mu}$\\
 & & 10$^{-2}$ keV & & erg cm s$^{-1}$ & \multicolumn{3}{c}{10$^{-13}$ erg cm$^{-2}$ s$^{-1}$} & & & ($\chi^2$/d.o.f.)\\
\hline
P1 & $2.80$  & $10.4\pm0.3$ & $4.5^{+0.5}_{-1.3}$ & $970^{+120}_{-440}$ & $33.2^{+4.2}_{-2.7}$ & $105.2^{+2.2}_{-11.2}$ & $58.3^{+7.6}_{-17.4}$ & 0.27 & 0.09 & 992/973\\
P2 & $2.63$  & $9.0\pm0.3$  & $4.6^{+0.6}_{-0.8}$ & $320^{+210}_{-110}$ & $35.3^{+2.4}_{-7.6}$ & $33.9^{+0.8}_{-1.1}$ & $33.0^{+4.6}_{-4.3}$ & 0.31 & 0.18 & 844/719\\
P3 & $2.69$  & $10.0^{+0.3}_{-0.4}$  & $3.1^{+0.7}_{-0.5}$ & $610^{+400}_{-140}$ & $34.8^{+3.2}_{-4.8}$ & $74.1^{+10.4}_{-7.8}$ & $36.0^{+6.0}_{-6.1}$ & 0.25 & 0.08 & 720/719\\
P4 & $2.63$  & $9.5\pm0.3$  & $3.1^{+0.8}_{-0.5}$  & $500^{+210}_{-220}$ & $34.1^{+7.3}_{-4.2}$ & $37.7^{+6.7}_{-1.3}$ & $25.9^{+7.6}_{-3.1}$ & 0.29 & 0.12 & 627/592 \\
P5 & $2.80$  & $7.8^{+0.9}_{-0.5}$  & $5.2^{+0.5}_{-1.3}$  & $970^{+120}_{-680}$ & $22.1^{+7.2}_{-2.0}$ & $45.6^{+1.6}_{-8.5}$ & $35.6^{+11.4}_{-9.2}$ & 0.21 & 0.23 & 682/683\\
P6 & $2.63$  & $9.5^{+0.3}_{-0.4}$ & $3.6\pm1.0$ & $520^{+220}_{-30}$ & $24.5^{+1.3}_{-4.7}$ & $37.8^{+6.9}_{-1.3}$ & $22.9^{+8.9}_{-3.5}$ & 0.24 & 0.14 & 722/767 \\
P7 & $2.71$  & $9.9\pm0.2$ & $3.2\pm0.3$ & $510^{+80}_{-90}$ & $49.8^{+4.1}_{-5.6}$ & $60.2^{+8.4}_{-6.3}$ & $46.9^{+11.2}_{-9.0}$ & 0.32 & 0.12 & 921/859\\
P8 & $2.63$  & $9.2^{+0.4}_{-0.5}$ & $3.6^{+1.7}_{-0.4}$ & $220^{+100}_{-110}$ & $41.0^{+3.6}_{-1.3}$ & $20.8^{+3.5}_{-1.7}$ & $29.6^{+13.2}_{-5.3}$ & 0.34 & 0.25 & 382/346\\
P9 & $2.63$  & $7.6^{+0.9}_{-1.0}$ & $8.6^{+0.9}_{-0.4}$ & $500^{+210}_{-310}$ & $11.4\pm1.4$ & $5.6^{+0.4}_{-0.5}$ & $16.8^{+5.5}_{-3.6}$ & 0.48 & 0.27 & 386/337\\
P10 & $2.63$  & $8.3\pm0.4$ & $7.0^{+0.5}_{-0.7}$ & $500^{+130}_{-270}$ & $32.1^{+1.3}_{-0.8}$ & $17.2^{+0.6}_{-0.8}$ & $36.7^{+5.5}_{-5.2}$ & 0.51 & 0.17 & 702/588\\
P11 & $2.63$  & $8.6^{+0.3}_{-0.2}$ & $4.8^{+0.5}_{-0.4}$ & $200^{+30}_{-60}$ & $42.9^{+1.4}_{-2.3}$ & $18.0\pm0.9$ & $37.4^{+5.2}_{-3.0}$ & 0.38 & 0.3 & 811/631\\
P12 & $2.63$  & $7.5^{+0.8}_{-0.7}$ & $5.7^{+1.8}_{-1.0}$ & $110^{+80}_{-50}$ & $24.3\pm5.1$ & $6.2^{+1.0}_{-1.5}$ & $24.2^{+8.3}_{-4.4}$ & 0.42 & 0.38 & 210/205 \\
P13 & $2.80$  & $10.0\pm0.2$ & $3.6^{+0.5}_{-0.6}$ & $1000^{+50}_{-340}$ & $53.5^{+4.0}_{-2.5}$ & $138.2^{+2.2}_{-6.1}$ & $84.9^{+7.9}_{-4.9}$ & 0.28 & 0.1 & 1241/1087 \\
P14 & $2.63$  & $9.2\pm0.2$ & $5.1^{+0.6}_{-0.7}$ & $490^{+20}_{-280}$ & $31.0^{+2.5}_{-2.1}$ & $25.9^{+0.7}_{-0.9}$ & $31.0^{+2.8}_{-4.2}$ & 0.31 & 0.23 & 815/704\\
P15 & $2.63$  & $8.9\pm0.3$ & $5.4^{+0.5}_{-0.3}$ & $500^{+240}_{-230}$ & $28.9^{+1.2}_{-0.6}$ & $14.8^{+0.6}_{-0.7}$ & $27.1^{+2.8}_{-3.0}$ & 0.32 & 0.33 & 649/577\\
P16 & $2.63$  & $8.6^{+0.4}_{-0.5}$ & $6.0\pm0.7$ & $500^{+110}_{-290}$ & $21.9\pm1.9$ & $12.6^{+0.6}_{-0.8}$ & $25.6^{+2.2}_{-3.1}$ & 0.30 & 0.37 & 477/471 \\
\hline\hline\\
\end{tabular}
\end{table*}

\subsubsection{Normalized Light Curves} \label{subsubsec:colour}

\begin{figure*}
\begin{center}
\leavevmode \epsfxsize=17cm \epsfbox{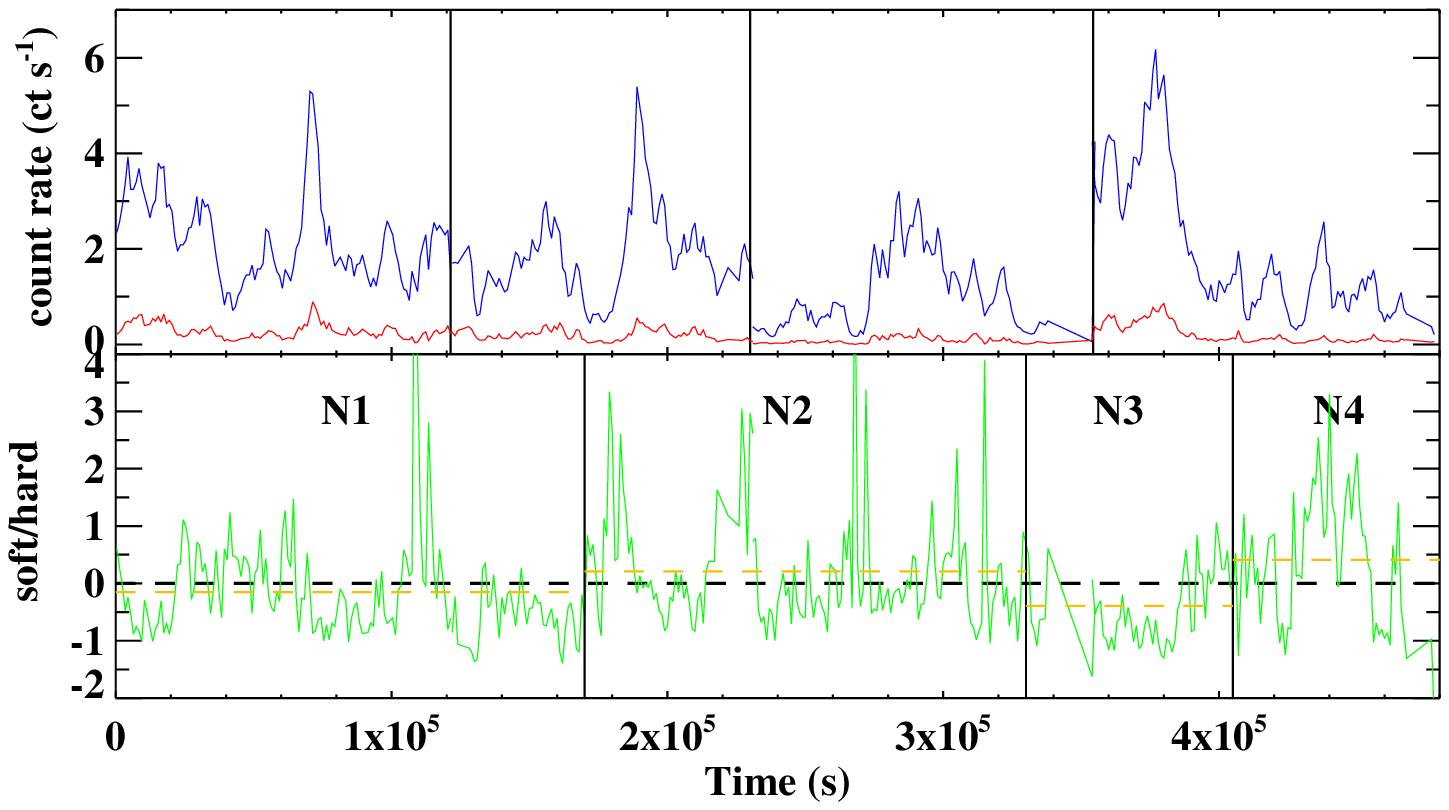}
\end{center}
\caption{The figure shows normalized soft (0.3-1.0 keV; blue), hard (1.2-5.0
keV; red) and colour(soft/hard; green) light curves. The orange dash line
represents the averaged value of the colour light curve in each segment. The time bin used in this plot is 1 ks.} \label{lc_norm}
\end{figure*}

Our previous analyses examine how the source behaves between different flux 
states. We now extract the 0.3-1.0 keV (soft) and 1.2-5.0 keV (hard) PN light 
curves with 10-ks time bins to see how the hardness ratio (hard/soft) evolves 
during the observation. However, the count rates of the soft light curve are
much larger than those of the hard light curve, so the hardness ratio remains 
small (slightly above 0; see upper panel in Fig. \ref{lc_norm}). 
In order to clearly visualise the evolution of hardness ratio, we normalised 
(subtracted the mean and divided by the standard deviation of the data) the
soft (blue) and hard (red) light curves in Fig. \ref{lc_norm}. As some values of
the soft light curve are also quite close to 0, the hardness ratio obtained from
original light curves hits infinity even after normalisation. We instead calculated
the colour (soft/hard) light curve and normalised it (lower panel in Fig.
\ref{lc_norm}). The gaps between the four observations have been discarded.

Based on the behaviour of the normalised hardness ratio, it can be seen that the 
observation can be roughly separated into four different periods. We defined 
period 1 (N1) to be 0-175 ks, period 2 (N2) to be 175-330 ks, period 3 (N3) to
be 330-405 ks, and period 4 (N4) to be 405-500 ks. In N1 and N3, the averages of soft/hard
ratios (orange dash line in lower panel of Fig. \ref{lc_norm}) fall below 0, implying the source is generally harder during these periods,
and softer during N2 and N4 (see also the orange dash lines in Fig.
\ref{lc_norm}). These periods are roughly commensurate with those used in the
time-dependent reverberation analysis presented by \cite{Kara13IRAS}, allowing
for direct comparison of the results obtained. For each of these periods, we
extracted spectra from each of the EPIC detectors, and applied the model used
previously. As before, parameters that are not expected to be time variable, e.g., 
$N_{\rm H}$, $a^{*}$, $i$, and $A_{\rm Fe}$, etc., are fixed at the best fit values 
from our time-averaged analysis, and we again use a simpler powerlaw form for 
the emissivity profile. The best-fit values for the remaining variable parameters
are listed in Table \ref{ta_colour} for each period. 

As expected based on our previous analysis, the flux of the powerlaw increases
and $\Gamma$ steepens during N1 and N3 (the higher flux periods). It is again
clear that $F_{\rm RDC}$ shows less variability than $F_{\rm PLC}$, and is
actually higher than $F_{\rm PLC}$ during N2 and N4. The inner emissivity
profile appears to be steeper during N2 and N4, which is consistent with results
of the flux-resolved and 30-ks time-resolved analyses. Again there is a hint that
the ionisation parameter of the higher ionisation reflection component ($\xi_{1}$)
correlates with the flux of the powerlaw component, but the values of N1, N2 and
N4 are comparable and consistent within the 90 per cent confidence. In principle
both the emissivity profile and disc ionisation parameter can give information of
the position and size of the corona, and we will discuss this in detail in section
\ref{sec_corona}.

\begin{table*}
 \caption{The table lists parameters of the best-fitting model of different periods 
which were selected using the normalised light curves (see Fig. \ref{lc_norm}).
Fluxes are again given in $10^{-13}$ erg cm$^{-2}$ s$^{-1}$, and $\xi$ in erg
cm s$^{-1}$. Parameters that are fixed at the same values with Table \ref{ta_ta}
are not listed here (see text).}
\label{ta_colour}
\centering
\begin{tabular}{@{}llcccc}
\hline\hline
Component & Parameter & \multicolumn{4}{c}{Value}\\
 & & N1 & N2 & N3 & N4\\
\hline
BBODY & $kT$ (10$^{-2}$ keV) & $9.6\pm0.2$ & $9.2^{+0.1}_{-0.2}$ & $9.6\pm0.2$ & $8.8\pm0.2$\\
 & Norm & $3.9^{+0.3}_{-0.4}\times 10^{-5}$ & $(4.0\pm0.1)\times 10^{-5}$ & $4.9^{+0.6}_{-0.2}\times 10^{-5}$ & $(3.0\pm0.1)\times 10^{-5}$ \\
 & $F_{\rm BB}$ & $32.4^{+2.2}_{-2.7}$ & $33.4^{+0.9}_{-0.4}$ & $41.3^{+3.2}_{-2.1}$ & $25.2^{+0.8}_{-0.4}$ \\
POWERLAW & $\Gamma$ & $2.69$ & $2.63$ & $2.80$ & $2.63$ \\
 & Norm & $(5.3\pm0.1)\times 10^{-4}$ & $(2.2\pm0.1)\times 10^{-4}$ & $7.4^{+0.2}_{-0.1}\times 10^{-4}$ & $(1.4\pm0.1)\times 10^{-4}$ \\
 & $F_{\rm PLC}$ & $59.3^{+2.0}_{-3.4}$  & $23.4\pm0.4$ & $93.7^{+1.6}_{-7.3}$ &  $14.5^{+0.3}_{-0.5}$\\
RELCONV & $q$ & $3.8^{+0.4}_{-0.5}$ & $4.9\pm0.2$ & $4.1^{+0.4}_{-0.6}$ & $5.8^{+0.3}_{-0.4}$\\
% & $q_{2}$ & $3.4\pm0.3$ & $3.6^{+0.2}_{-0.3}$ & $3.8^{+0.3}_{-0.4}$ & $4.0\pm0.4$\\
% & $R_{\rm break}$ ($R_{\rm g}$) & $2.0^{+0.2}_{-0.1}$ & $2.0^{+0.2}_{-0.1}$ & $2.0^{+0.2}_{-0.1}$ & $1.9^{+0.3}_{-0.1}$\\
EXTENDX &  $\xi_{1}$ & $530\pm70$ & $500^{+10}_{-210}$ & $830^{+190}_{-250}$ & $500^{+100}_{-220}$ \\
 & ${\rm Norm}_{1}$ & $1.4^{+0.3}_{-0.2}\times 10^{-8}$ & $1.5^{+1.1}_{-0.2}\times 10^{-8}$ & $1.5^{+0.7}_{-0.3}\times 10^{-8}$ & $1.0^{+0.7}_{-0.1}\times 10^{-8}$ \\
 & $\mathcal{R}_{1}$ & $\sim0.27$ & $\sim0.43$ & $\sim 0.30$ & $\sim0.32$\\
% & $\xi_{2}$ & $20\pm2$ & $20^{+1}_{-2}$ & $22^{+2}_{-1}$ & $20^{+1}_{-6}$ \\
 & ${\rm Norm}_{2}$ & $4.6^{+1.5}_{-1.1}\times 10^{-6}$ & $3.7^{+0.2}_{-0.3}\times 10^{-6}$ & $9.9^{+1.0}_{-3.0}\times 10^{-6}$ & $5.6^{+0.2}_{-0.4}\times 10^{-6}$ \\
 & $\mathcal{R}_{2}$ & $\sim0.11$ & $\sim0.13$ & $\sim0.13$ & $\sim 0.33$\\
 & $F_{\rm RDC}$ & $36.5^{+5.8}_{-5.0}$ & $29.8^{+0.6}_{-2.0}$ & $69.9^{+6.2}_{-12.2}$ & $27.2^{+2.8}_{-2.0}$ \\
$\chi^{2}/d.o.f.$ &  & 1646/1475 & 1570/1225 & 1426/1194 & 1027/863 \\
\hline\hline\\
\end{tabular}
\end{table*}

\section{Discussion} \label{discussion}

We have presented various analyses of the spectral properties and in particular
the strong spectral variability exhibited by IRAS13224-3809 during the long 
500\,ks \textit{XMM-Newton} observation performed in 2011. The results
obtained are well-explained by the relativistic reflection model initially
presented in \cite{Ponti10} and \cite{Fabian13}, and we confirm that
IRAS13224-3809 hosts an almost maximally-rotating black hole. The reflected
emission shows less variability than the powerlaw continuum, which is expected
if gravitational light-bending effects are important. The model also includes a
low temperature blackbody, which we find roughly follows the $L \propto T^{4}$
relation expected for stable thermal emission, suggesting this has a physical
origin and is not just a phenomenological component included in the model to
obtain good fits. We will further discuss the possible origin of this component in
section \ref{sec_bb}. Our time-resolved analysis allows us to investigate how the
spectral properties evolve between different flux levels, and we also discuss
below how the source geometry may change by comparing results obtained
from our spectral analyses and previous timing analyses (\citealt{Kara13IRAS}).

\subsection{General Trends}

In both the flux-resolved and time-resolved analyses we presented above, it is
clear that the reflected emission shows less variability than the powerlaw
continuum. $F_{\rm PLC}$ can vary by a factor of more than 10 during the entire
observation (Table \ref{ta_tr}), while $F_{\rm RDC}$ only changes by a factor of
$\sim$5. This agrees with the prediction of the gravitational light-bending
model for a dynamic/geometrically evolving corona located in the region of
strong gravity close to the black hole. \citet{Miniutti04} showed that, for a
simple lamppost disc--corona geometry, if the variability observed from the
intrinsic X-ray source is related to changes in its proximity to the black hole, the
reflection component should remain relatively constant due to gravitational light
bending effects, as with a greater proximity, more of the X-ray emission is bent
towards the black hole and lost over the event horizon rather than `seen' by the
disc. The same phenomena has been observed in a number of other Type 1
AGN (e.g. \citealt{Vaughan04, Fabian12}),
and also in the Galactic black hole binary XTE\,J1650-500 (e.g. \citealt{Rossi05,
Reis13}). In this scenario, since the reflection component should show less
variability than the powerlaw continuum, when the observed flux from the
source decreases it should progressively become more reflection-dominated,
consistent with our spectral decomposition. 

Assuming a compact, isotropic source above the accretion disc, the emissivity
of the reflected emission from the disc would naturally decrease with increasing
radius. In the simplest case without considering any relativistic effects, an
emissivity index of $\sim$3 is expected \citep{Reynolds97}. However, in
general relativity, if the X-ray source is close to the black hole, photons from
the X-ray source are bent towards the central black hole and inner disc. As a
result, the closer the source is to the black hole, the more the inner
disc is preferentially illuminated, resulting in steeper emissivity profiles. More
details of how relativistic effects modify the emissivity profile around AGN can
be found in \citet{Wilkins12}, but the basic expectation is that increasingly
reflection-dominated states should be accompanied by increasingly steep
emissivity indices. Again, our spectral decomposition returns exactly this result;
as shown in Fig. \ref{parameter} (panels c and d), the emissivity index becomes
steeper as $F_{\rm RDC}/F_{\rm PLC}$ increases and as the observed flux of
the powerlaw continuum decreases.

\subsection{The Blackbody Component}
\label{sec_bb}

A blackbody component is sometimes used to fit the spectrum of AGN
phenomenologically to explain the soft excess, but this component is not
necessarily generated by the accretion disc. It is common to see the blackbody
emission from the accretion disc in the X-ray spectrum of stellar-mass black
holes (e.g. \citealt{Remillard06rev, Done07rev}).  However in AGN, due to a
much larger black hole mass, the accretion disc is cool and the blackbody
component usually peaks in the EUV band with a temperature of a few eV.  For
AGN with a very massive central black hole, the inclusion of a blackbody
component when fitting an X-ray spectrum is a way to model a soft excess of
unknown origin. However, for AGN with relatively small central black holes and
high accretion rates (e.g. NLS1s), the disc blackbody component can extend
into the soft X-ray band.  As an extreme NLS1, it is possible that the soft X-ray
blackbody component of IRAS13224-3809 originates from the accretion disc.

Another possibility is that the inner part of the accretion disc is irradiated by a
combination of direct coronal emission and returning radiation, heating it
sufficiently to cause a blackbody component to appear in the soft X-ray band. 
Strong gravitational light-bending effects take place when the spin $a^{*}$
$>$ 0.9 causing some of the radiation from the accretion disc (thermal
photons, reflection, etc.) to return to the disc, resulting in modifications in the
spectrum (\citealt{Cunningham76}). Since IRAS 13224-3809 has a high spin,
effects caused by returning radiation may be important. If the accretion disc is
heated by either coronal photons or reflected photons then the blackbody
component should be driven by the powerlaw or reflection component. In Fig.
6f, it appears that $F_{\rm BB}$ is correlated to both $F_{\rm PLC}$ and
$F_{\rm RDC}$, though stronger with the latter.

Returning radiation causes multiple X-ray reflection (\citealt{Ross02}) which
creates additional spectral components in the final X-ray spectrum. The soft
X-ray spectrum is steeper and featureless and the K-shell emission/absorption
features of iron strengthen. \cite{Ross02} demonstrated that multiple reflection
could enhance the flux in the soft X-ray band, producing results similar to
those caused by a blackbody component.

\subsection{The Black Hole Mass} \label{subsec:mass}

With knowledge of the distance and inclination of IRAS13224-3809, we can
attempt to deduce the mass of the central black hole from the blackbody
emission, given this appears to follow $L\propto T^{4}$ relation, implying a
constant emitting area. Since the emission comes from the region affected by
strong gravity, relativistic effects should be considered. The {\sevensize BBODY}
model gives the colour temperature of the blackbody component. A spectral
hardening factor $f_{\rm col}$ should be included to obtain the effective
temperature ($T_{\rm eff}\times f_{\rm col}=T_{\rm col}$). A typical
$f_{\rm col}$ of 1.7 has been commonly used \citep{fcol} and works well on
stellar-mass black holes in the disc dominated state. In the case of NLS1s,
however, the colour temperature correction could be as large as $\sim$2.4
\citep{Ross92, Done12}. The energy flux released from the accretion disc is
related to the central black hole mass $M_{\rm BH}$. By measuring the disc flux,
with the appropriate relativistic corrections, $M_{\rm BH}$ can be obtained.
\citet{kerrbb} introduced the {\sevensize KERRBB} model, which calculates the
emission from a multi-temperature, steady state, general relativistic accretion
disc around a Kerr black hole. We replaced the {\sevensize BBODY} component
with {\sevensize KERRBB} model and re-fit the time-averaged spectrum to
estimate the black hole mass. The inclination, spin parameter and distance have
been fixed to the best-fitting values of the time-averaged analysis (Table \ref{ta_ta}, 
broken powerlaw emissivity) for consistency. Assuming $f_{\rm col} $ = 2.4, we obtained a
mass of $M_{\rm BH} = 3.5^{+5.5}_{-0.6} \times 10^{6} M_{\odot}$ and an effective mass
accretion rate of $2.74^{+0.14}_{-0.38}\times10^{23}$g s$^{-1}$, which imply the $L/L_{\rm Edd}$
to be $\sim$0.23. The {\sevensize KERRBB} gives a flux of $\sim2.6\times10^{-17}$ erg cm$^{-2}$ s$^{-1}$ \AA$^{-1}$
at 2310\AA, which significantly under-predicts the OM UVM2 flux $3.07\times10^{-15}$ 
erg cm$^{-2}$ s$^{-1}$ \AA$^{-1}$. The best-fitting
black hole mass is smaller than the expected value $\sim 10^{7} M_{\odot}$ obtained from the
reverberation analysis (\citealt{Kara13IRAS}), though the upper limit
is close. The $\sim$90 s soft lag detected
in IRAS13224-3809 implies that the source should be roughly three times more
massive than 1H0707-495, which harbours a black hole of $M_{\rm BH} \sim
7\times10^{6}M_{\odot}$. 

The discrepancy in the black hole mass between spectral and timing analyses may indicate 
the origin of the blackbody component. Note that the {\sevensize KERRBB} 
model uses the Novikov-Thorne profile for calculation. If the blackbody 
component is not induced by emission coming directly from the accretion disc,
the black hole mass obtained from {\sevensize KERRBB} is not valid and other 
method should be used for mass estimation. It is possible that
the blackbody component is produced by irradiation or returning radiation mentioned
in section \ref{sec_bb}. Although these alternatives cannot be examined
by current models, the much smaller black hole mass obtained from spectral analysis
and the underestimation of optical flux, may be hints of irradiation or returning radiation.

%As most power of corona is directed to a small region, the reflection might be not correctly modelled and caused the underestimation.

%The blackbody radiation emitting area can be calculated using the relation $L=A\sigma T^{4}$, where $A$ is surface area and $\sigma$ the Stefan-Boltzmann constant. The calculation indicated that the emitting area is roughly from 1.25 $R_{\rm g}$ to 7 $R_{\rm g}$ (assuming $M_{\rm BH}=1.5\times 10^{6} M_{\odot}$). This again shows that most energy is released from a small area near the central black hole. Since the black hole mass is lower than expected, the outer edge of emitting area, 7 $R_{\rm g}$, should be an upper limit.

\subsection{The Size of the Corona}
\label{sec_corona}

\citet{Kara13IRAS} discovered that the reverberation lag from the soft excess has
a larger amplitude (by a factor of $\sim$3) during high-flux states, and occurs
on slightly longer timescales. This can be explained if the corona becomes more
extended during high-flux states. Furthermore, in the case of the Galactic
binary XTE\,J1650-500, \cite{Reis13} also argue the characteristic size of the
corona that is changing, through study of the simultaneous evolution of the
reflected emission and the quasi-periodic oscillations exhibited by the coronal
emission. Since the reverberation lag is relatively insensitive for a corona
expanding radially with a constant vertical extent \citep{Wilkins13}, the result
implied that the corona is more vertically-extended when brighter. The
high-flux states in \citet{Kara13IRAS} are comparable with the periods N1 and
N3 defined in section \ref{subsubsec:colour}, and the low-flux states similar
with N2 and N4, although we note that N1 has a much longer exposure than the
low flux period examined by \citet{Kara13IRAS}.

In principle, it may also be possible to determine the scale height of the corona
from the ionisation of the accretion disc, which is defined as $\xi =
L_{\rm ion}/nR^{2}$, where $L_{\rm ion}$ is the ionising luminosity, $n$ is the
hydrogen number density of the disc, and $R$ is the distance between the
source and the disc. Under the assumption that $n$ does not vary substantially
within a short period of time, changes in $\xi$ imply either changes in the
height of the corona, or changes in the intrinsic ionising luminosity from the
corona. If the corona is more vertically-extended during high-flux states, and
$L_{\rm ion}$ remains constant (i.e. the observed variability is driven by changes
in source geometry rather than intrinsic luminosity, as suggested above),
one might naively expect that $\xi$ should be lower for N1 and N3, and higher
for N2 and N4, given the change in the characteristic distance between the
inner disc and the corona. Instead, the ionisation parameter of the first (higher
ionisation) reflection component seems to be positively correlated with the
observed powerlaw flux\footnote{The ionisation parameter of the second (lower
ionisation) reflection component ($\xi_{2}$) was consistent with being constant
throughout the observation. However, though our analysis assumes the two
reflection components are co-spatial in a radial sense (as the same relativistic
convolution is applied to each), the simplest picture is that the higher ionisation
component is a hotter skin on the surface of the accretion disc, and thus we
would probably expect this component to respond more than the lower
ionisation component regardless of the nature of the variability.} (although
$\xi_{1}$ for N1 is not significantly higher than that of N2 and N4, a positive
correlation is also implied by the analysis presented in section \ref{subsec:flux},
which covers a broader range of flux states).

However, this expectation does not fully account for the effects of gravitational
lightbending, as it does not take into account the accitional continuum flux lost
to the black hole. The ionisation parameter can be re-written as $\xi =
F_{\rm incident}/n$, where $F_{\rm incident}$ is the incident flux as seen by
the inner disc. If the albedo of the disc remains constant, changes in $F_{\rm
incident}$ are actually best probed by changes in the reflected flux $F_{\rm RDC}$. Thus,
based on the predictions of the gravitational lightbending model, one should in
fact expect that even if the intrinsic ionising luminosity from the corona is
constant, the ionisation of the disc should be higher when the flux is higher, as
in the regime of strong gravitational lightbending the reflected flux and the
observed powerlaw flux should be positively correlated (zone 1 in Figure 2 of 
\citealt{Miniutti04}). Given the positive correlation between $F_{\rm PLC}$ and
$F_{\rm RDC}$ observed, and the apparent positive correlation between $\xi_1$
and $F_{\rm PLC}$, these results would seem to be consistent with the corona
being more extended when the flux is brighter.

Additionally, the evolution of the emissivity index can also help to constrain
the evolution of the corona. As discussed previously, the emissivity index is
steeper when the source is fainter and more reflection-dominated (see Figure
\ref{parameter}). This also indicates the corona is likely becoming more compact
in the lower flux periods, resulting in stronger lightbending. Focusing again on
the periods highlighted in section \ref{subsubsec:colour} for direct comparison
with \cite{Kara13IRAS}, we again see evidence for the same trend (Table
\ref{ta_colour}), albeit over a smaller dynamic range than probed with the
analyses presented in sections \ref{subsec:flux} and \ref{subsubsec:30ks}. We
also tried replacing the {\sevensize RELCONV} convolution kernel with the
{\sevensize RELCONV\_LP} version (\citealt{relconv_lp}), which assumes a
lamp-post geometry for the corona, i.e. a point source located at some height
$H$ above the spin axis of the black hole. Rather than fit the emissivity profile
with a simple parameterisation, this model allows the height of the point source
to be fit directly, and self-consistently computes the emissivity profile for that
geometry. 

As expected, applying this to the hardness-selected spectra (N1-4; section
\ref{subsubsec:colour}) we find that the height obtained for the corona is larger
in the higher flux states (N2, N4) than in the lower flux states (N1, N3), evolving
from $\sim$2\,$R_{\rm g}$ to $\sim$3\,$R_{\rm g}$. Making the same
replacement for the flux-resolved spectra (F1-4; section \ref{subsec:flux}), we
see the same trend, with the corona height systematically evolving from
$\sim$2\,$R_{\rm g}$ in the lowest flux state to $\sim$4\,$R_{\rm g}$ in the
highest. Although the trend is exactly as expected for a corona that is larger in
higher flux states, the range inferred for $H$ from {\sevensize RELCONV\_LP}
is slightly smaller than inferred from the reverberation analysis
(\citealt{Kara13IRAS}). However, the point source geometry assumed by
{\sevensize RELCONV\_LP} is an obvious over-simplification, particularly for the
scenario suggested in \cite{Kara13IRAS}, in which the corona has some extent
and its characteristic size evolves with flux. Therefore, even though there is
some tension between the exact quantitative evolution of the corona between
the spectral results obtained with {\sevensize RELCONV\_LP} and the
reverberation analysis, the fact that both imply the same general evolution is
very encouraging for this interpretation.

\section{Conclusions}

We have presented a detailed analysis of the spectral properties, and in
particular the spectral variability exhibited by the extremely variable NLS1
IRAS13224-3809 during the long \textit{XMM-Newton} observation obtained
in 2011. For the latter, we investigate how the spectrum evolves by examining
different states selected on both flux and a spectral hardness, as well as
undertaking a systematic time-resolved spectral analysis. These spectra are
interpreted in the context of the well established relativistic disc reflection
model, based on recent works by \cite{Ponti10} and \cite{Fabian13}. We find
that the reflected emission from the disc is much less variable than the
powerlaw continuum, and as the source flux drops, the spectrum becomes
progressively more reflection-dominated. Furthermore, as the source becomes
more reflection-dominated, the emissivity index simultaneously steepens.
These trends are exactly as expected if the variability is dominated by a
compact, centrally located X-ray corona which contracts as the source flux
drops, resulting in increased gravitational lightbending which in turn produces
more reflection-dominated states, and increases the preferential illumination
of the innermost accretion disc. This is also broadly consistent with the
observed evolution of the X-ray reverberation properties (\citealt{Kara13IRAS}),
which strongly supports this interpretation.

Finally, the best-fit model also includes a contribution from a very soft
blackbody. We find this appears to follow the expected $L \propto T^{4}$
relation for thermal emission from a stable emitting region, supporting a
physical (rather than phenomenological) origin for this emission component.
Given the high spin obtained for IRAS13224-3809 and the clear influence of
gravitational lightbending, this may be related to heating of the inner disc by
both radiation from the corona, and from `returning' radiation originating
emitted/reflected by the disc which is bent back towards the disc surface.

\section*{Acknowledgements}
This work was greatly expedited thanks to the help of Jeremy Sanders in
optimizing the various convolution models. We thank our referee, Chris Done, 
for helpful comments.

\bibliographystyle{mn2e_uw}
\bibliography{iras13224}

\label{lastpage}
\end{document}